\documentclass[aps, prd, twocolumn, superscriptaddress, floatfix, showpacs,
               nofootinbib]{revtex4}

\usepackage{amsmath}
\usepackage{amssymb}

\usepackage{graphicx}
\usepackage{subfig}

\usepackage{color}

\usepackage{natbib}

\newcommand{\specialcell}[2][c]{%
  \begin{tabular}[#1]{@{}c@{}}#2\end{tabular}}

\newcommand{\f}{f}
\newcommand{\fdot}{\dot{\f}}
\newcommand{\fddot}{\ddot{\f}}
\newcommand{\To}{T_{\textrm{obs}}}
\newcommand{\mGRID}{m}
\newcommand{\mumin}{\mu_{\mathrm{min}}}
\newcommand{\muminAvg}{\langle\mumin\rangle}

\newcommand{\dcc}{LIGO-P1400207-v1+}

\begin{document}

\title{The effect of timing noise on targeted and
    narrow-band coherent searches for continuous gravitational waves
    from pulsars}

    \author{G. Ashton}
    \email[E-mail: ]{G.Ashton@soton.ac.uk}
    \author{D.I. Jones}
    \affiliation{School of Mathematics,
                 University of Southampton, 
                 Southampton SO17 1BJ}
    \author{R. Prix}
    \affiliation{Max Planck Institut f{\"u}r Gravitationsphysik 
                 (Albert Einstein Institut),
                 30161 Hannover, Germany}

\date{\today} 

\begin{abstract} 

Most searches for continuous gravitational-waves from pulsars use Taylor expansions in the
phase to model the spin-down of neutron stars.  Studies of pulsars
demonstrate that their electromagnetic (EM) emissions  suffer from
\emph{timing noise}, small deviations in the phase from Taylor expansion
models. How the mechanism producing EM emission is related to any
continuous gravitational-wave (CW) emission is unknown; if they either
interact or are locked in phase then the CW will also experience timing
noise. Any disparity between the signal and the search template used in
matched filtering methods will result in a loss of signal-to-noise ratio
(SNR), referred to as `mismatch'. In this work we assume the CW suffers a
similar level of timing noise to its EM counterpart.  We inject and recover
fake CW signals, which include timing noise generated from observational
data on the Crab pulsar.  Measuring the mismatch over durations of
order~$\sim 10$~months, the effect is for the most part found to be small.
This suggests recent so-called `narrow-band' searches which placed upper
limits on the signals from the Crab and Vela pulsars will not be
significantly affected. At a fixed observation time, we find the mismatch
depends upon the observation epoch.  Considering the averaged mismatch as a
function of observation time, we find that it increases as a power law with
time, and so may become relevant in long baseline searches.

\end{abstract} 

\pacs{04.80.Nn, 97.60.Jd, 04.30.Db}
\maketitle

\section{Introduction}
\label{sec: narrow-band introduction} 
Rotating neutron stars capable of supporting non-axisymmetric mass
distributions will emit continuous gravitational
waves\footnote{Note that in general ``continuous waves'' can
    refer to any quasi-monochromatic long-lasting gravitational-wave
        signals, such as emitted by binaries of white dwarfs, neutron
        stars or black holes, which would be detectable by LISA or pulsar
        timing arrays. Here we refer to CWs exclusively in the context of
        spinning nonaxisymmetric neutron stars as relevant to ground-based
        detectors.}
(CWs) due to their
time-varying quadrupole moments. These may be detectable by next-generation ground-based detectors. The emitted
signals can persist for longer than
typical search durations, but are weak in amplitude, making them difficult to detect
in the noise of the detector.

To find a signal, CW searches use matched
filtering techniques such as the $\mathcal{F}$-statistic \citep{Jaranowski1998}
which compare the output of the detector with a template.  These techniques are
powerful provided that the signal and template remain coherent for the duration
of the observation. If the signal can be perfectly matched by a template then
the signal to noise ratio, used to quantify the detection likelihood, scales as
$\rho^{2} \propto \To$ (e.g.\ see~\citep{Prix2009}). This suggests searching
over longer observations increases the chances of making a detection.

The templates must model the monotonic spin-down of the source due to the
electromagnetic (EM) and gravitational torque; this is done by Taylor expanding
the phase:
\begin{equation}
\phi(t) = \phi_{0} + 2\pi\left(\f_{0} (t - t_{0}) + 
          \frac{\fdot_{0}}{2!} (t - t_{0})^{2} 
           %+ \frac{\fddot}{3!} (t - t_{i})^{3} 
           \right)
           + \ldots
           \,,
\label{eqn: Taylor}
\end{equation}
where $t_{0}$ is the reference time at which the pulsar frequency, and
spin-down parameters $[\phi_{0}, \f_{0}, \fdot_{0}, \ldots]$ are defined. Note
that all times refer to the solar system barycentre and we assume the 
timing model has already correctly accounted for the dispersion measure, proper motion
and other parameters as discussed in \citet{Edwards2006}.
Pulsar astronomers fit this model to observed time of arrivals (TOAs). If the
best fit model is accurate enough to track the pulsar to within a single
rotation the resulting timing solution is described as \emph{phase-connected}.
Often such solutions are capable of tracking the pulsar over durations greater
than  a year \citep{Lyne2012book}.  For gravitational-wave searches, this level of accuracy
motivates the use of the same Taylor expansion phase models to account for the
spin-down.  Pulsar observers measure the frequency $\f_{0}^{EM}$ and higher
order coefficients describing the rotation of the pulsar itself. In this
work we will consider searches for
emission from non-axisymmetric neutron stars at
$\f_{0}^{CW}=2\f_{0}^{EM}$~\citep{Shapiro83}; from hereon all frequencies
and spin-downs refer to the pulsars CW emission.

While Taylor expansion models are on average reliable enough to track the
spin-down, pulsars do show deviations. This can either be in the form of
glitches, occasional sudden increases in the rotation frequency, or 
continuous low-frequency 
variations known as \emph{timing noise}. Glitches and timing-noise may be
related phenomena, but are distinguishable by their relative frequency
and magnitudes. In this work we restrict our focus to timing noise (we will comment
        on glitches in sec.~\ref{sec: Minimum mismatch as a function of the observation epoch})

Timing noise is
often represented by structure in the timing residual, which is the difference
between the best fit Taylor expansion, typically up to second order in spin-down
$\ddot{f}_0$, and the observed phase. Timing noise refers specifically to
deviations from Taylor expansions that are intrinsic to the pulsar and not to
systematic errors such as dispersion in the interstellar medium.
\citet{Hobbs2010} conducted a wide ranging study on timing noise across the
pulsar population.  They concluded, amongst other things, that timing noise is
ubiquitous and inversely correlated to the age of the pulsar.  There already
exists measures used to quantify the strength of timing noise such as the
$\Delta_{8}$ value introduced by \citet{Arzoumanian1994}, the generalisation
of the Allan variance \citep{Matsakis1997}, the covariance function of the
residuals \citep{Coles2011}, and fitting for timing noise as part of the pulsar
timing model \citep{Lentati2014}. These do not convert directly 
into the effect that timing noise may have on CW searches for pulsars. To quantify this, we
need to measure the \emph{mismatch} due to timing noise. This is closely
related to the loss of signal to noise ratio due to the imperfect matching
between the template and signal (which we define explicitly in section
\ref{sec: narrow-band method}).

Although a variety of models exist to interpret timing noise \citep{Lyne2010, Cordes1981}, there is
currently no consensus on a single mechanism. However, for the issue of timing
noise and CW searches, we only need to consider the relation between the
components of the neutron star which produce the EM and CW signals.  This was
investigated by \citet{Jones2004} who identified three possible scenarios.
First, the two signals are strongly coupled: the same timing noise will be
observed in both. Second, the two signals are loosely coupled: a similar, but
different level of timing noise will be observed in both. Third, timing noise
exists only in the EM signal, there is no corresponding variations in the CW
signal. Of course these are really three cases from a full spectrum of
possibilities which could also include the pulsars CW signal being significantly more noisy than its EM signal.

The significance of timing noise will vary between different types of pulsar CW
searches; these can be divided into targeted, narrow-band, directed, and
all-sky searches.  \emph{Targeted} searches involve a single known pulsar where
an estimate of the spin parameters has been obtained from the EM signal. If we
assume that the EM and CW signals are strongly coupled, then we can use a
\emph{single-template} targeted search. Under this assumption, when the level
of timing noise in the EM signal is small, then a single Taylor expansion is
sufficient.  If instead the level of timing noise is large, then the EM data
can be used to account for it; this is done by applying an adapted
matched-filtering phase-model that closely follows the observed EM phase model
\citep{Pitkin2004}. If instead we assume that the EM and CW signals are
loosely coupled, then we should perform a \emph{narrow-band} search in a small
area of parameter space.  These narrow-band searches aim to allow for small
frequency offsets between the EM and CW signals, such as could be caused by
free precession, or a finite coupling time between the two components of the
neutron star \citep{LIGO2008}. \emph{Directed} searches look for non-pulsing
neutron stars predicted by other means such as at the centre of the super-nova
remnant Cassiopeia A. An \emph{all-sky} search involves searching over the
entire sky for unknown pulsars. For both directed and all-sky searches the lack
of EM data necessitates wide bands in the frequency and its derivatives. For
fully coherent matched filtering methods these searches can rapidly become
computationally prohibitive. To circumvent this, semi-coherent search
techniques are used that incoherently combine short fully-coherent sections of
data \citep{LIGO2012}; these will be less sensitive to timing noise.
Nevertheless, semi-coherent searches  ultimately need to be followed up by
targeted fully coherent searches, for which timing noise may be an issue.

For the properties of the CW signal, the most general case is that it will
exhibit some timing noise, but it could be different to the timing noise
observed in the EM signal. Until a detection is made, we can only make
assumptions about how the two are correlated. To probe these assumptions, we
will define two special cases corresponding to different sorts of errors in a
CW search:
\begin{itemize}

\item {\bf Special Case 1:} Timing noise, exactly like that in the EM signal,
exists in the CW signal but is not included in the template. This will result
in a loss of signal to noise ratio for searches which assumed that timing noise
was negligible. The error potentially affects the narrow-band, directed, and
all-sky searches since the level of timing noise is unknown. The single
template targeted searches will not be effected since they either check that
the level of timing noise is negligible, or correct for it using an adaptive
phase model.

\item {\bf Special Case 2:} Timing noise is included in the template but does
    not exist in the signal. This will result in a loss of signal to noise
    ratio for single-template targeted searches that account for timing noise
    using an adapted phase model (for example \citet{LIGO2008}).  Instead,
    these searches will now erroneously introduce timing noise into the
    template while the signal will be a smooth Taylor expansion. 

\end{itemize}

In this work we will mimic narrow-band and single-template searches to directly
simulate special case 1. Specifically, we will inject a fake CW signal
which contains a realisation of timing noise, and recover it using
templates based on a single global Taylor series. This tests the scenarios in
which the timing noise in the CW signal is either exactly coupled to the EM
signal, or they are at least similar.  However, this also quantifies special
case 2 since the signal and template are interchangeable in matched filtering
methods.  That is, timing noise in the signal but not in the template is
equivalent to timing noise in the template but not the signal.

While all known pulsars are potential CW sources, young pulsars are the most
promising due to their large spin-downs (see~\citet{LIGO2010} for a
        review). However, it was found by \citet{Hobbs2010} that the amount of timing
noise is correlated with the spin-down  magnitude. This motivated us to study
the effect of timing noise on CW searches for neutron stars with large spin-downs.

The realisation of timing noise we will use to investigate timing noise
in CW pulsar sources is based on the young Crab pulsar. The Crab is a
potentially detectable  source of gravitational waves due to its high spin-down
rate and it has the highest spin-down upper limit compared to the LIGO noise
floor \citep{LIGO2008}. The EM signal from the Crab is well documented (see
        sec.~\ref{sec: timing noise as described by the crab ephemeris}) and contains
exceptional levels of timing noise: it was estimated by \citet{Jones2004} that
such levels of timing noise in the CW signal may cause an issue for current
searches.

Several targeted searches have already
been performed for CWs from the Crab pulsar.
A single-template search for CWs from the Crab pulsar was performed on data
collected during the LIGO S5 science run \citep{LIGO2008}.  This search used
the Crab ephemeris and an adapted phase model to account for timing noise.  In
addition to this single-template search, a narrow-band search for signals from
the Crab was also performed by \citet{LIGO2008} on the S5 data. Another
narrow-band search for the Crab was carried out using data from the VIRGO
VSR4 science run along with a search for the Vela pulsar \citep{LIGO2015}. 

The structure of this paper is as follows. In section \ref{sec: timing noise as
described by the crab ephemeris} we describe the observational data available
from the Crab ephemeris and discuss its relation to CW searches. In section
\ref{sec: narrow-band method} we describe the signal injection and recovery
method.  Results from this method are presented in section \ref{sec:
narrow-band results}: we begin by considering the effect timing noise has on
narrow-band searches, then we consider the mismatch on stretches of data for
which narrow-band searches have been performed; we further investigate how the
mismatch depends upon epoch; and finally examine how the mismatch depends on
the duration of observation. We summarise our results in section \ref{sec:
narrow-band conclusions}.

\section{Timing noise as described by the Crab ephemeris}
\label{sec: timing noise as described by the crab ephemeris}
The monthly Crab ephemeris \citep{Lyne1993} provides the phase evolution of the
EM signal between 1982 and the present and can be found at
\url{http://www.jb.man.ac.uk/pulsar/crab.html}. It is unlike most timing data
for pulsars where a timing model consists of the model parameters (position, spin-down,
        etc.) given at a single reference time.
For the Crab ephemeris, each monthly update consists
of the frequency and spin-down coefficients along with a reference time
coinciding with the TOA of a pulse at the solar system barycentre. The
coefficients are calculated by  least-squares fitting of a Taylor expansion to
the TOAs. The reference time for each month is chosen as the TOA of
the pulse closest to the mid-point; this is done to minimise the average phase
error of the local Taylor expansion.  The period of a month is short enough
such that these coefficients and equation~\eqref{eqn: Taylor} track the
rotational phase during the month.

\begin{figure}[htb]
\centering
    \includegraphics{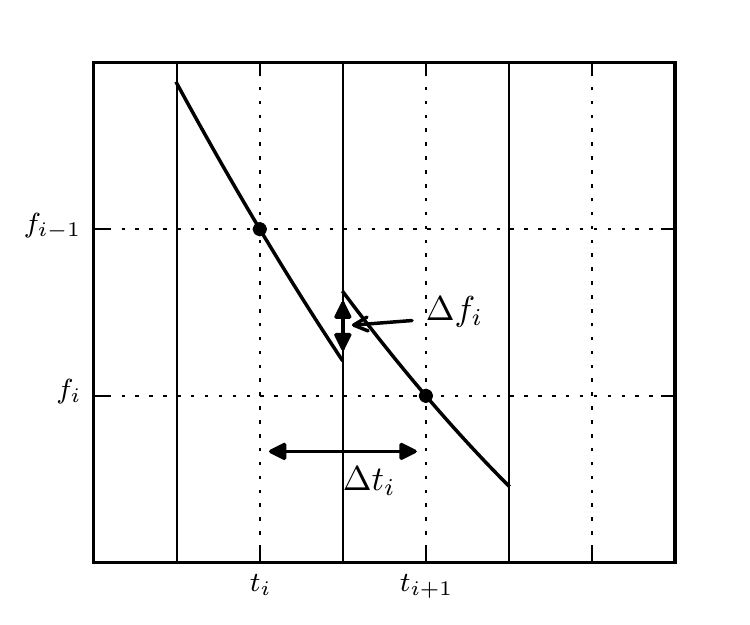} 
\caption{Illustration of the jumps between 'local' per-month templates in 
    frequency space, defining the frequency jump $\Delta \f$. This 
        depiction amplifies the order of magnitude of $\Delta \f$ in order to
        highlight the timing-noise: for the jumps in the ephemeris, $\Delta \f$
        is several orders of magnitude smaller than the change in frequency
        due to spin-down alone.}
\label{fig: template jumps}
\end{figure}

The Crab ephemeris gives a distinct picture of the variations due to timing
noise superimposed on the monotonic spin-down. To illustrate how this manifests
itself, figure~\ref{fig: template jumps} depicts the frequency evolution in two
adjacent months. Notice that a discontinuity occurs at the interface between
months.  Such discontinuities will occur in the spin-down, frequency, and
phase; timing noise can then be described by the magnitude of these jumps.
From the Crab ephemeris it can be shown that the distribution of jumps in
phase, frequency and spin-down appear to follow standard normal distributions.
This is consistent with timing noise models consisting of a large number of
small unresolved events accumulating over a month (e.g. the models considered
by \citet{Cordes1981}).

Timing noise is usually depicted by structure in the phase residuals calculated
by removing the best fit Taylor expansion to the phase from the real phase
evolution. A best fit Taylor expansion consists of a single set of coefficients
$\f_{0}$, $\fdot_{0}$, and $\fddot_{0}$ valid over the \emph{entire}
observation period. To make this distinct from the \emph{local} Taylor
expansions describing the evolution in each month this will be referred to as
the \emph{global template}.  In figure~\ref{fig: template jumps} we see that if
the discontinuity is non-zero, then it is impossible for any global Taylor
expansion template to exactly match the local templates in both months. The
phase residual, and hence timing noise, results from the inability to match a
single global template to all the local ones.

In this work we aim to quantify the significance of timing
noise in CW searches by generating signals from the Crab ephemeris. This
is an empirical description of timing noise and so we
make no assumptions on the underlying astrophysical model.

\section{Method}
\label{sec: narrow-band method}
We now describe the method to quantify the effect of timing noise on CW
searches for signals from isolated pulsar. 
To be relevant to current pulsar CW search methods, we will base our
method on the narrow-band searches of \citet{LIGO2008} and \citet{LIGO2015}. The results can be interpreted as measuring the consequence of
special case 1 on narrow-band and single-template searches; that is we assume
the CW signal has a similar level of timing noise as the EM signal and search
using global Taylor expansion templates. For this study, a single-template
search refers to a single Taylor expansion template and not the adapted phase
model proposed by \citet{Pitkin2004}. 

We begin by generating a CW signal emulating timing noise using the Crab
ephemeris. This is done by stringing together month-long smooth Taylor expansion
signals. Each month uses the corresponding month from the Crab
ephemeris for the Taylor expansion coefficients. The `jumps' at the interface
between months constitutes the timing noise.
In more detail:
\begin{enumerate}

\item From the ephemeris select a period of data consisting of the reference
times $t_i$, frequency $\f_i$, and spin-down
$\fdot_i$ for each month $i$.

\item \label{fit} Generate the phase as a function of time from the data and
then fit a global Taylor expansion up to $\ddot{f}$ for the whole
observation time.  The fit results in set of interpolated coefficients
$[\f_{0}, \fdot_{0}, \fddot_{0}]$ at a global reference time halfway through
the data. These coefficients are used to centre the narrow-band search
parameters.

\item We supplement the local monthly data $\{t_i, \f_{i}, \fdot_{i}\}$ with
the fixed value of $\fddot_{0}$ calculated in the previous step. The phase of
the CW signal is always zero at each monthly reference time $t_i$ of the
ephemeris, which by construction coincides with a pulse arrival time.

\end{enumerate}

In this process we have assumed that a fixed value of $\fddot_{0}$ is
sufficient. This can be justified by considering the next term in the Taylor
expansion \eqref{eqn: Taylor} and typical values of $\dddot{\f} \sim 10^{-30}$
Hz/s$^{3}$.  Over typical search durations~$\sim 1$yr this term contributes
less than a radian to the phase, and it can therefore be safely neglected.

We use the \verb+LALSuite+ \cite{lalsuite} gravitational-wave analysis routines
to generate a fake CW signal; for these experiments we work
without any simulated detector noise. The standard tool to generate fake CW
signals uses single Taylor expansion models. Therefore, to include timing noise
in the signal we do the following.

\begin{enumerate}
\setcounter{enumi}{3}

\item We inject each month-long Taylor expansion generated from the Crab ephemeris
lasting for only the duration of that month.
This
method creates a fake CW signal, lasting several months, which includes timing noise corresponding to the monthly
ephemeris.

\end{enumerate} 

Once we have produced data, we then use \verb+LALSuite+ tools to recover the
signal using the $\mathcal{F}$-statistic \citep{Jaranowski1998}. This is a 
matched filtering method in which the output of the detector is compared to a
signal template (see \citet{Prix2009} for more details).

Two types of searches are performed: a single
template search at the interpolated coefficients $[\f_{0}, \fdot_{0},
\fddot_{0}]$ and a narrow-band search in $\f$ and $\fdot$ centred on the
interpolated coefficients. These searches were found to be sufficient
to find the signal to within a reasonable mismatch, so more sophisticated
methods where not required.

The narrow-band consists of a grid of points in
$f$ and $\fdot$. As found by \citet{LIGO2008} we find searching over
$\fddot_{0}$ to be unnecessary for this experiment and so it is kept fixed using
the value found in step 2 above.
 The grid spacing is parameterised
by $\mGRID$, the one-dimensional maximal mismatch between two adjacent Taylor
expansion templates. From
\citet{LIGO2013_EAH} the corresponding grid spacing is given by 
\begin{align}
    d\f = \frac{\sqrt{12 \mGRID}}{\pi \To} &&&
    d\fdot = \frac{\sqrt{720 \mGRID}}{\pi \To^{2}},
\label{eqn: Grid spacing}
\end{align}
where $\To$ labels the observation time.

For the single-template and at each grid point in the narrow-band search, we
measure the squared SNR value~$\rho^{2}$.  In order to quantify the relative
loss compared to the perfectly phase-matched squared SNR
$\rho^{2}_{\mathrm{s}}$, we define the mismatch in the usual way (e.g.\ see
\citet{Prix2007}) as
\begin{equation}
    \mu = \frac{{\rho^{2}_{\mathrm{s}} - 
                                  \rho^{2}}}{\rho^{2}_{\mathrm{s}}}.
\label{eqn:mismatch}
\end{equation}
It is well known (e.g.\ see \citet{Prix2009}) that the SNR for a
perfectly phase-matched signal is independent of the signal phase
evolution. Therefore, in the absence of timing
noise the measured value of~$\rho^{2}$ can reach the maximum value 
of~$\rho^{2}_{s}$, and the mismatch therefore vanishes in that template.
In the presence of timing noise, even the best-matching template will
suffer some mismatch, and this effect will increase with the level of
timing noise.

In the single-template search, we measure a single mismatch value.  The
single-template search can also be interpreted to quantify the error made in
special case 2, when the template is adapted to account for EM timing noise but
none exists in the CW signal.  We can think of the narrow-band search as
repeating the single-template search over a grid of points; this allows us two
degrees of freedom, corresponding to the frequency and spin-down parameters,
over which to minimise the mismatch. The grid point with the minimum mismatch,
which we denote by $\mumin$, is the best candidate and will used to quantify
the success of the search.  Because the narrow-band can minimise the mismatch,
$\mumin$ must always be equal or smaller than the mismatch in the
single-template search.

\section{Results}
\label{sec: narrow-band results}
\subsection{The effect of timing noise on narrow-band searches} We begin by
describing how timing noise degrades a narrow-band search. This is done by
comparing the result for a signal containing no timing noise with a signal
generated from the Crab ephemeris between MJD 45150 and 56668. This period
holds no special significance and is used simply to demonstrate the essential
features of a signal containing timing noise.

\begin{figure}[htb]
\centering
\includegraphics{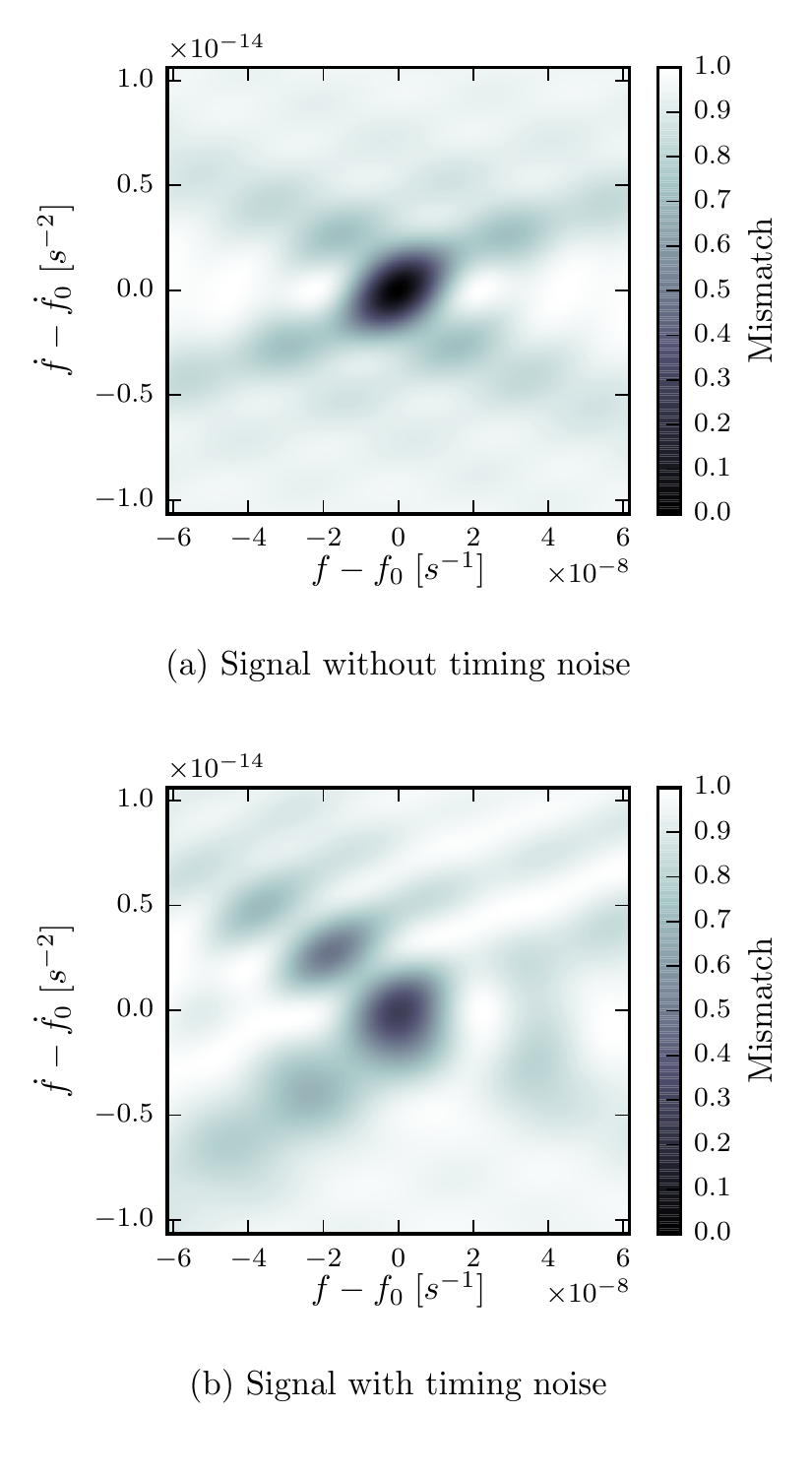}
\caption{In figure (a) we show the mismatch as a function of parameter space
    for a signal without timing noise. The injected signal has
    parameters~$[f_{0}, \dot{f}_{0}, \fddot_{0}]$, as a result the mismatch has
    a minimum at this point. This can be compared with figure (b) showing the
mismatch from a signal including timing noise. The signal is generated from the
Crab ephemeris between MJD 45150 and 45668.}
\label{fig: narrow-band example}
\end{figure}

In figure ~\ref{fig: narrow-band example} we show the mismatch as a function of
parameter space offset for (a) a signal without timing noise, and (b) a signal
containing timing noise. The signal without timing noise is injected at the
interpolated coefficients~$[f_{0}, \fdot_{0}, \fddot_{0}]$.  Therefore, we find
the minimum mismatch with $\mumin=0$ at exactly the centre of the grid and the
iso-mismatch contours in the local neighbourhood around the origin are well
described by ellipses (e.g.\ see \citet{Prix2007}).  For the signal with timing
noise (b) we notice two distinctive effects: the minimum achievable mismatch
$\mumin$ is non-zero, and the iso-mismatch contours around the best-match
template are more irregular and less well described by ellipses.  In the
following we will quantify the effect of timing noise by considering only the
location and value of the minimum mismatch grid point in the narrow-band
search.

\subsection{Results relevant to recent narrow-band searches}

First we consider two particular periods of the Crab ephemeris corresponding to
recent narrow-band searches for the Crab: the LIGO S5 period \citep{LIGO2008}
and the VIRGO VSR4 period \citep{LIGO2015}. The mismatch in the single-template
and the minimum mismatch for the narrow-band searches during both periods are
listed in table~\ref{tab: Results}. For these periods timing noise is
found to produce a mismatch of~$\approx 1\%$. As expected, the narrow-band
mismatch is smaller than the single-template search. The fractional difference
between the two searches is relatively small. 

Provided that the timing noise observed in the CW signal is at the same level
(or less) as that observed in the EM signal, this result signifies that the
recent LIGO and VIRGO narrow-band searches would not suffer significantly
from the effects of timing noise.

In addition to producing a mismatch, timing noise may result in the best
candidate being found at some distance from the centre of the narrow-band
search.  However, we find that the distance from the centre of the grid is
small when compared to the grid spacing used in actual narrow-band searches
such as the S5 and VSR4.  For the S5 period narrow-band search, we find that
the minimum mismatch was a fraction~$\sim 0.01$ of the grid spacing used in the
\citet{LIGO2008} search.  At the resolutions used in real narrow-band searches,
the effects of timing noise on the location of the minimum mismatch will not be
evident.

\begingroup
\small
\begin{table}[ht] 
\centering
\begin{tabular}{lccc} 
    & 
    \specialcell{Dates \\ MJD} & 
    \specialcell{Single  template \\ $\mu$} &
    \specialcell{Narrow band \\ $\mumin$} \\ \hline
S5 & 53673 - 53977 & $0.00968$ & $0.00933$ \\ 
VSR4 & 55681 - 55839 & $0.00659$ & $0.00584$ \\ 
\end{tabular}•
\caption{Measurements of the mismatch during the S5 and VSR4 narrow-band search
         periods.}
\label{tab: Results}
\end{table}
\endgroup

\begin{figure}[htb]
\centering
\includegraphics{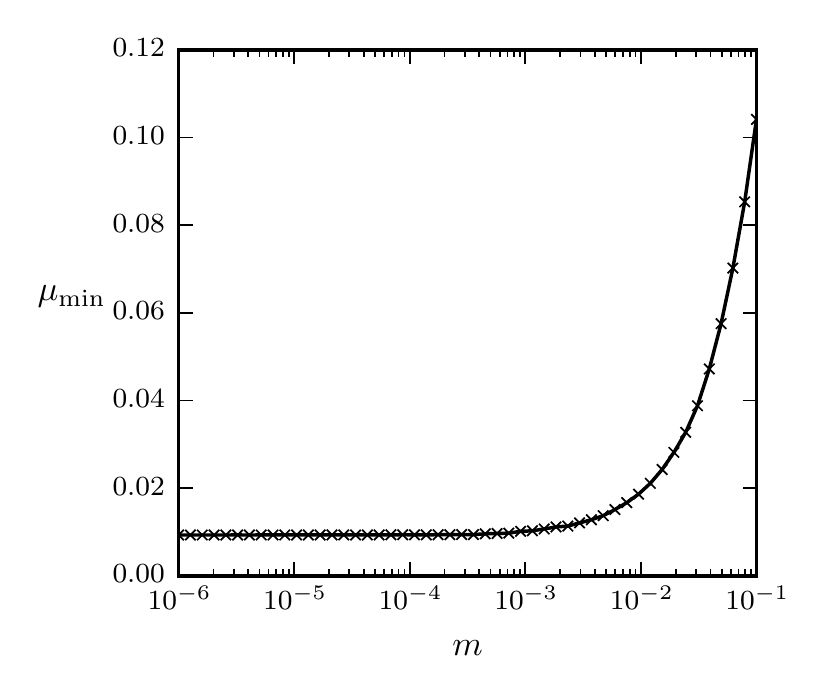}
\caption{Measured best mismatch $\mumin$ as a function
of grid spacing parameter $\mGRID$ (see eqn.~\eqref{eqn: Grid spacing}), for the Crab pulsar over the S5 period.
    This demonstrates that $\mumin$ plateaus at a nonzero mismatch suggesting
    we are resolving a mismatch due to timing noise instead of the effect of finite
    grid resolution.}
\label{fig: conv}
\end{figure}

Figure~\ref{fig: conv} shows the convergence of the measured best mismatch
$\mumin$ for the narrow-band search over the S5 period with the value
of~$\mGRID$.  This demonstrates that the non-zero values of $\mumin$ given in
table~\ref{tab: Results} are not the result of grid coarseness.  For signals
without timing noise, the measured best mismatch $\mumin$ will have a minimum
of $\sim\mGRID$ when the putative signal is located halfway between grid
points. In the limit of~$\mGRID \rightarrow 0$ we then expect the measured
mismatch to tend to zero. Instead, for a signal with timing noise we observe a
plateau after some initial reduction. This indicates that the grid is now
\emph{fully} resolving the variations due to timing noise.
•

\subsection{Minimum mismatch as a function of the observation epoch}
\label{sec: Minimum mismatch as a function of the observation epoch}

We will now investigate how the best mismatch $\mumin$ varies as a function of
the observation epoch. We only show the narrow-band search, as the results were
found to be very similar for the single-template search. The method consists of
measuring the mismatch $\mumin$ in a 6-month window, which is shifted in 1
month intervals over all the available ephemeris data. The observation time of
6 months is chosen to be similar to typical CW search durations. We are
restricted to multiples of 1 month by the frequency of updates to the Crab
ephemeris.

Timing noise is not the only variability in the spin-down of pulsars - they can
also undergo sudden increases in rotation frequency known as \emph{glitches}.
The Crab frequently glitches and these are catalogued by \citet{Espinoza2011}
and available at \url{http://www.jb.man.ac.uk/pulsar/glitches.html}.
The mechanism which causes a glitch is not well understood and may involve
unpredictable variations in the CW signal. As a result, targeted CW searches
either avoid periods with known glitches \citep{LIGO2008}, or 
allow for an arbitrary jump in gravitational wave phase at the time of
the glitch \citep{LIGO2010}. For this work, we ignore
the complicating factor introduced by glitches and consider 
only the effect of timing noise. We do this by
omitting windows which include glitches from the search by using
the aforementioned glitch catalogue.

We begin by searching in a small
$40\times40$ grid in frequency and spin-down, with a fixed grid space
mismatch of~$\mGRID=1\times10^{-5}$, and the grid spacing as defined in
equation~\eqref{eqn: Grid spacing}. It is possible that the minimum mismatch is
found at the edge of the narrow-band grid; such candidates  are not true local
minima in the mismatch. If this is the case, the search is repeated with an
increasingly larger grid size, but the same fixed grid spacing.  This process
continues until we find a minimum mismatch which is not at the edge of the
grid. 

\begin{figure}[htb]
\centering
\includegraphics{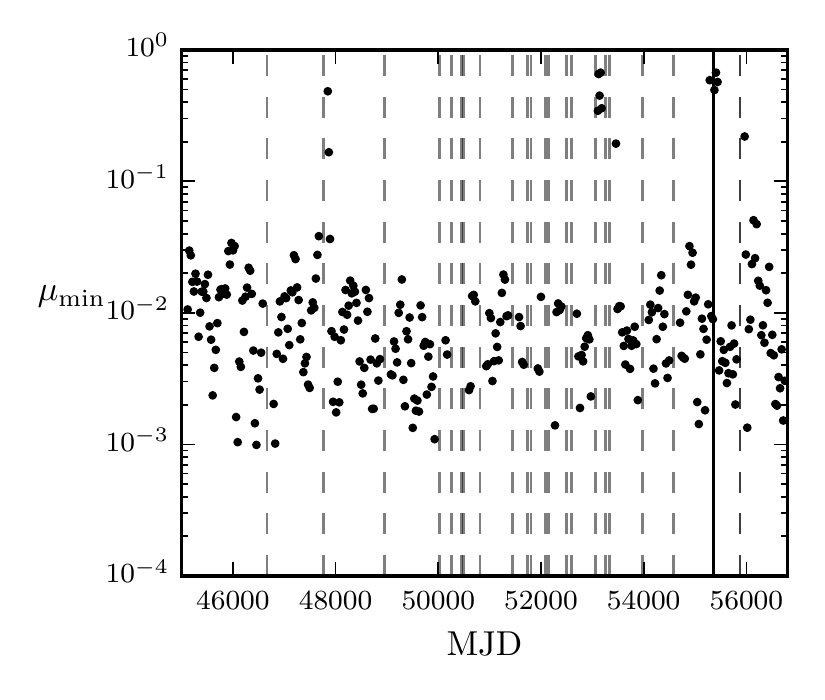}
\caption{Minimum mismatch $\mumin$ found in 6-month sliding window searches
    as a function of epoch
     at the centre of  window.  Vertical dashed lines
    indicate glitch events as described by \citet{Espinoza2011}. The solid  
    vertical line indicates the date MJD 55362, a period of anomalously 
    large mismatch.}
\label{fig: sliding window}
\end{figure}

Figure~\ref{fig: sliding window} shows the measured minimum mismatch in the
narrow-band search for a sliding 6-month window at the centre of the
observation time. The mismatch due to timing noise is the low level noise
occurring in between glitches. Greater mismatches are observed in the
post-glitch periods; this is expected as the relaxation time after glitches for
the Crab is of the order 1 month \citep{Lyne2012book}.  We note the presence of
an anomalous period of large mismatch for all  windows that include the
ephemeris time MJD 55362. The cause for this is unclear from the available
data, but it may be caused either by a measurement error or a small undetected
glitch. In general, we find that the level of mismatch due to timing noise is
between $\mumin \sim 10^{-3} - 10^{-2}$ for these 6-month searches.

•

\subsection{Averaged minimum mismatch as a function of the observation duration}
\label{sec: averaged mismatch as a function of the observation duration}

We can study the averaged behaviour of the mismatch $\mumin$ as a function of
time by varying the size of the sliding window in the previous section. This
was done for both the narrow-band and single-template searches; the mismatch
from the narrow-band search was found to be  a fraction~$\lesssim 0.1$ smaller
on average than the single-template search. We therefore will only present
results from the narrow-band search. The shortest possible window~$\sim 6$
months is restricted by the number of points needed to generate a fit to the
phase.  Setting the upper limit at~$\sim 17$ months retains a statistically
meaningful number of points to average over. Having obtained the data from all
sliding window sizes in this range we want to analyse the average behaviour as
a function of the observation time. Before doing this we filter results in the
following ways:

\begin{itemize}
    \item We do not consider any windows that
    include or are bounded by glitch events

    \item Windows including the anomalous epoch MJD 55362 are omitted. We 
    wish to study the fluctuations due to timing noise, and this
    period is either an unidentified glitch, or another highly unusual and
    unrepresentative form of timing noise
        
   \item While each entry of the ephemeris is on average valid over a whole
   month, some months were truncated due to glitches. The sliding window, which
   works on a fixed number of entries of the ephemeris will occasionally be
   shorter than average. To ensure we are averaging over windows of a
   similar length we omit windows for which the observation time differs from
   the average by 2 weeks.
   
\end{itemize}

\begin{figure}[ht]
\centering
\includegraphics{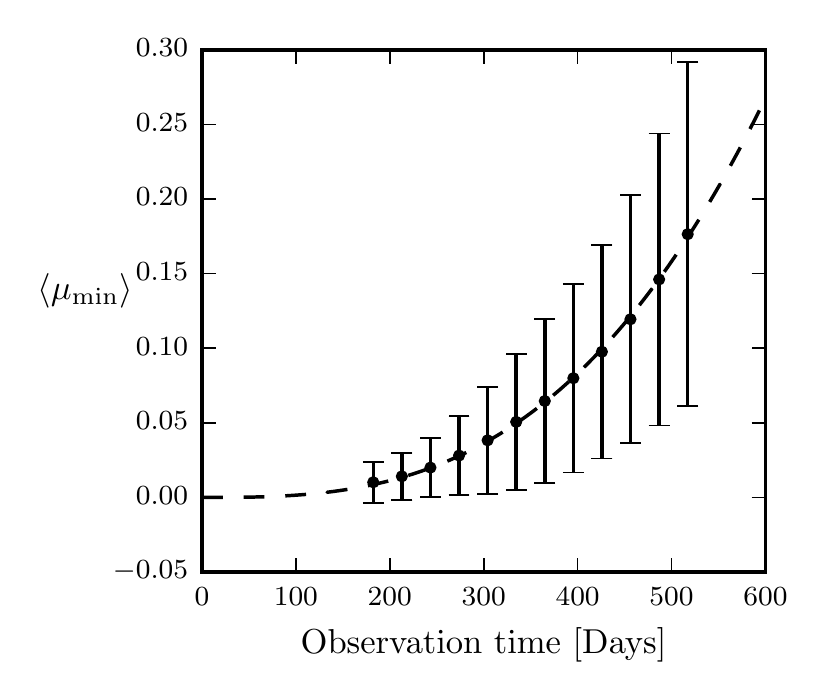}
\caption{Averaging the mismatch for sliding window searches and varying the
observation times. The points give the mean while the bars correspond to one 
standard deviation.}
\label{fig: mismatch Tobs}
\end{figure}

In figure~\ref{fig: mismatch Tobs} we plot the averaged minimum
mismatch $\muminAvg$ as a function of observation time.
This indicates a growth of $\muminAvg$ with observation time
resembling a power law.
•

To quantify the growth of the mismatch, we perform a least-squares fitting 
to a power law. Fitting the expression:
\begin{equation}
\muminAvg_{\textrm{fit}} = \kappa 
                                   \left(\frac{\To}{1\textrm{ sec}}\right)^{n},
\end{equation}
we find the best fit parameters
\begin{align}
    \kappa&=1.5 \pm 0.8 \times 10^{-23} \\ 
n&=2.88 \pm 0.030 
% Tmax_estimate = 592.09227117 
% At which <m> = 0.25747842399
% S5 mismatch = 0.0362443664997
.
    \label{eqn: fit values}
\end{align}

For perfectly matched signals the squared SNR
increases linearly \citep{Prix2009} with observation time.
This suggests that longer observation
times yield a greater likelihood of detection. The power law fit with~$n > 1$
implies that the average mismatch from
Crab timing noise grows faster than the squared SNR.
Gains in
SNR from longer observation time will therefore eventually be
outweighed by the increasing mismatch from timing noise.
To estimate when this may occur, we can rearrange equation~\eqref{eqn:mismatch}
to give
\begin{equation}
    \rho^{2} = \rho^{2}_{\mathrm{s}}\left(1 - \mumin\right).
\end{equation}
Substituting the time dependencies for the perfectly matched SNR and the
averaged mismatch we have
\begin{equation}
    \rho^{2} \propto \To - \kappa \To^{n+1}.
\end{equation}
Differentiating and solving for~$\To$ yields an expression for the
observation time (in seconds)
beyond which the~$\rho^{2}$ value of a signal containing timing 
noise starts to decrease
\begin{equation}
\To = \left(\frac{1}{\kappa(1+n)}\right)^{1/n}.
\end{equation}
For the fit values from equation~\eqref{eqn: fit values}, this yields
a critical observation time of
$\To\approx 600$~days after which the mismatch exceeds
$\muminAvg \approx 0.25$.
% These results can be found in the FitResultsNarrowBand.tex file
In this case it is no longer true that further increases in
observation time will yield greater detectability.

\citet{Jones2004} estimated the maximum time the signal and template would
remain coherent given a random walk in  frequency. A crude method used a
phase residual of 1 rad for the decoherence criteria. For the Crab, this
estimates the decoherence time at 200 days. We can improve upon this result by
setting a mismatch of~$0.1$ as the decoherence criteria; using the fit to the
averaged mismatch this gives us a decoherence time of $\To\approx 400 $~days.

    The growth of mismatch as a power law is suggestive of
random walk timing noise models (see~\citet{Cordes1981}) 
    for which the rms phase residual also grows as a power law. However,
    such a scaling in the phase residual must first be converted into a mismatch,
    which will depend on the search method, before a comparison can be made. In 
    future work we will present a method to achieve this.

%For the S5 narrow-band search which
%lasted 10 months, the averaged mismatch predicts a value of~$0.036$; this is
%relatively large when compared to table~\ref{tab: Results}. This suggests
%either the S5 period is relatively quite or the data is skewed by outliers
%close to glitches or in the noisy period of MJD 55362. For longer observation
%times these outliers become less important as the average mismatch increases.

\section{Conclusions}
\label{sec: narrow-band conclusions}
We have used observational data on the Crab pulsar to characterise the possible
effects of timing noise on coherent targeted single-template  and narrow-band
continuous gravitational-wave searches for pulsars.  This was done by generating fake
signals based on the Crab ephemeris data and searching for them using templates
without timing noise. Our analysis clarifies the impact for current searches;
accordingly, our methods mimic those used by \citet{LIGO2008} and \citet{LIGO2015}.

Our primary results is summarised by Fig.~\ref{fig: mismatch Tobs}: 
when considering the average mismatch as 
a function of observation time,  we find that the averaged
mismatch grows as a power law. In addition to this, we found two interesting
aspects when considering the data without averaging over the epoch:

Firstly, for the S5 and VSR4 narrow-band searches, if the timing noise in the
CW signal from the Crab is at a similar level (or lower) to that in the EM
signal, then we find it will only have a small~($\approx1\%$) effect on the
measured squared SNR of the putative signal.  We found the mismatch in
single-template searches to be only fractionally larger than the narrow-band
searches.  This also suggests phase-adapted searches would not be significantly
effected if the signal does not contain timing noise.

Secondly, searching over all available Crab data with a 6-month window, we
looked at the mismatch as a function of observation epoch. Post glitch periods
tend to admit significant levels of mismatch; this is expected due to the
exponential recovery from the glitch. (We also discovered a period around MJD
55362 which has a large mismatch and is not connected to a known glitch). The
narrow-band and single-template searches performed similarly in this and
subsequent tests. Typically the mismatch due to timing noise for 6-month
searches was found to be between $10^{-3}$ and $10^{-2}$.

The scope of this work can be extended to  directed and all-sky searches, which
target young rapidly spinning down stars which may emit the strongest CWs.
These stars are also known to exhibit the highest levels of timing noise and
glitch frequently. Crucially the lack of EM data means we cannot be certain a
glitch does not occur during the observation and we cannot account for timing
noise in the signal.  In future work we would like to quantify both these
effects and estimate safe upper limits for the search durations.
It would also be interesting to consider CW searches for low-mass X-ray binary systems.
These are believed to exhibit a stronger form of timing noise known as
``spin-wandering'', which constrains the maximal coherence time to
the order of a few days before it would lead to a complete loss of
SNR, thereby limiting the best achievable sensitivity
\cite{2014arXiv1412.0605T,2015arXiv150200914L,ScoX1:MDC1}.

\section{Acknowledgements}
GA acknowledges financial support from the University of Southampton and
the Albert Einstein Institute (Hannover).  DIJ acknowledge support from
STFC via grant number ST/H002359/1, and also travel support from NewCompStar
(a COST-funded Research Networking Programme).
All authors are grateful for useful feedback from
members of the Continuous Waves group of the LIGO Scientific Collaboration
and the Virgo Scientific Collaboration.
This paper has been assigned document number \dcc.

\bibliography{NarrowBandBibliography}
 
\end{document}